\pdfoutput=1

\documentclass[final]{elsart}

\usepackage{graphicx}
\usepackage{amsmath}
\usepackage{amssymb}
\usepackage{times}
\usepackage{verbatim}

\DeclareGraphicsExtensions{.pdf}

\usepackage{amssymb}

\begin{document}

\begin{frontmatter}



\title{Evolutionary Games on Networks and Payoff Invariance Under Replicator Dynamics}


\author[label1]{Leslie Luthi}
\author[label1]{Marco Tomassini}
\author[label1]{Enea Pestelacci}

\address[label1]{Information Systems Department, HEC, University of Lausanne,Switzerland}

\begin{abstract}
The commonly used accumulated payoff scheme is not invariant with respect to shifts of
payoff values when applied locally in degree-inhomogeneous population structures.
We propose a suitably modified payoff scheme and we show both formally and by numerical 
simulation, that it  leaves the replicator dynamics invariant with
respect to affine transformations of the game payoff matrix. We then show empirically that, using
the modified payoff scheme, an interesting amount of cooperation
can be reached in three paradigmatic non-cooperative two-person games in populations that are 
structured according to graphs that have a marked degree inhomogeneity, similar to actual graphs found in society. The three games are the Prisoner's Dilemma, the Hawks-Doves and the Stag-Hunt. This confirms previous important observations that, under certain conditions, cooperation may emerge in such network-structured populations, even 
though standard replicator dynamics for mixing populations
prescribes equilibria in which cooperation is totally absent in the Prisoner's Dilemma, and it is
less widespread in the other two games. 

\noindent \end{abstract}

\end{frontmatter}


\section{Introduction and Previous Work}
\label{intro}

Evolutionary game theory (EGT) is an attempt to study the conflicting objectives among agents playing non-cooperative
games by using Darwinian concepts related to frequency-dependent selection of strategies in a population~\cite{maynard82,weibull95,hofb-sigm-book-98}, instead of positing mathematically convenient but practically unrealistic conditions of agent rationality and common knowledge  as is customary in classical game theory~\cite{Myerson}. Two concepts play a prominent role in EGT: the first is
the idea of an \textit{evolutionarily stable strategy} (ESS) and the second is the set of equations representing the dynamical
system called \textit{replicator dynamics} (RD)~\cite{taylor-jonker}. Both concepts are related to an ideal situation in which there are
random independent encounters between pairs of anonymous memoryless players using a given strategy in an
infinite population. In such a situation, a strategy is said to be an ESS if a population using that
strategy  cannot be invaded by a small amount of mutant players using another strategy (this idea can be expressed in rigorous mathematical terms, see~\cite{weibull95}). However, the ESS concept has a static
character, i.e.~it can be applied only once the population has reached a robust rest point following certain dynamics.
In other words, an ESS is restricted to the analysis of a population in which all the members play the same
strategy and the stability of the strategy is gauged against the invasion of a small amount of individuals 
playing another strategy. 
The replicator dynamics, on the other hand, given an initial population in which each strategy is present with
some frequency, will end up in attractor states, as a result of the preferential
selection and replication of certain strategies with respect to others. Simply stated, strategies that do better than
the average will increase their share in the population, while those that do worse than the average will
decline. The link with standard game theory is the following: the ESSs for a game, if at least one exists,
is a subset of the game-theoretic equilibria called Nash equilibria (NE). The attractor states of the dynamics
may be fixed points, cyclical attractors, or even chaotic attractors in some situation. However, a
result of replicator dynamics guarantees that, among the rest points of the RD, one will find the NE and thus, a fortiori, the game's ESSs~\cite{weibull95}. These results pertain
to infinite populations under standard replicator dynamics; they are not necessarily true when the
assumptions are not the same e.g., finite populations with local interactions and discrete time evolution, which is the case considered here.

Several problems arise in EGT when going from very large to finite, or even small populations
which are, after all,  the normal state of affairs in real situations. For example, in small populations theoretical ESS might not be reached, as first observed by Fogel et al.~\cite{fogeletal97,fogeletal98} and Ficici et al.~\cite{ficici-pollack-07}, and see also~\cite{nowak-et-al-finite-04}. 
The  method affecting the selection step can also be a source of difference with
respect to standard EGT, even for infinite mixing populations. Recently, Ficici et al.~\cite{ficici-pollack-05} have  shown that using selection methods different from  payoff proportionate selection, such as truncation, tournament or ranking leads to results that do not converge to the  game theory equilibria postulated in standard replicator dynamics. 
Instead, they find different non-Nash attractors, and even cyclic and chaotic attractors. 

While the population structure assumed in EGT is panmictic, i.e.~any player can be chosen to interact with any
other player, it is clear that ``natural'' populations in the biological, ecological, and socio-economical realms often
do have a structure. This can be the case, for instance, for territorial animals, and it is even more common in human
interactions, where a given person is more likely to interact with a ``neighbor'', in the physical or relational
sense, rather than with somebody else that is more distant, physically or relationally. Accordingly, EGT concepts have
been extended to such structured populations, starting with the pioneering works of Axelrod~\cite{axe84} and Nowak and
May~\cite{nowakmay92} who used two-dimensional grids which are regular lattices. However, today it is becoming clear that regular lattices are only
approximations of the actual networks of interactions one finds in biology and society.  Indeed,
 it has become apparent that many real networks are neither regular nor random
graphs; instead, they have short diameters, like random graphs, but much higher clustering
coefficients than the latter, i.e.~agents are locally more densely connected. These networks are
collectively called \textit{small-world} networks (see~\cite{watts99,newman-03}).
Many technological, social, and biological networks are now known to be of this kind. Thus, 
research attention in EGT has recently shifted from mixing populations, random
graphs, and regular lattices towards better models of social
interaction structures~\cite{social-pd-kup-01,santos-pach-05,tom-luth-giac-06,luthi-pest-tom-physa08}. 

Fogel et al.~\cite{fogeletal97,fogeletal98} and Ficici et al.~\cite{ficici-pollack-05,ficici-pollack-07} studied the deviations that occur in EGT when some of the standard RD assumptions are not fully met.
In this paper we would like to address another problem which arises when using RD in network-structured populations.
In the standard setting, populations are panmictic, i.e.~any agent may interact with any other agent in the population. However,  in complex networks, players may have a widely different number
of neighbors, depending on the graph structure of the network interactions. On the other hand, panmictic populations
may be modeled as complete graphs, where each vertex (agent) has the same number of neighbors (degree).
The same is true for any regular graph, and thus for lattices, and also, at least in a statistical sense, for Erd\"os--R\'enyi random
graphs~\cite{bollobas-random}, which have a Poissonian degree distribution. In the cases where the number of neighbors is the same for all players, after each agent has played the 
game with all of its neighbors, one can either accumulate or average the payoff earned by a player
in order to apply the replicator dynamics. Either way, the result is the same except for a 
constant multiplicative factor. However, when the degrees of agents differ widely, these two ways of calculating
an agent's payoff give very  different results, as we show in this paper. Furthermore, we show that
when using accumulated payoff, the RD is not invariant with respect to a positive affine transformation of the payoff matrix as it is prescribed by the standard RD theory~\cite{weibull95}. In other words, the game depends on the particular payoff values and is non-generic~\cite{samuel97}. Finally, we propose another way of calculating an agent's payoff that both takes into account
the degree inhomogeneity of the network and leaves the RD invariant with respect to affine transformations of the payoff
matrix. We illustrate the mathematical ideas with numerical simulations of
three well-known games: the \textit{Prisoner's Dilemma}, the  \textit{Hawk-Dove},
and the \textit{Stag-Hunt} which are universal metaphors for
conflicting social interactions. 

In the following, we first briefly present the games used for the simulations. 
Next, we give a short account of the main population graph types used in this work, mainly for the sake of making
the paper self-contained. Then we describe the particular replicator dynamics that is used on networks, followed by an analysis of the influence of the network degree inhomogeneity on an individual's payoff calculation. The ensuing discussion of  the results of many numerical experiments should help illuminate the theoretical points and 
the proposed solutions. Finally, we give our conclusions.

\section{Three Symmetric Games}
\label{games}

The three representative games studied here are the Prisoner's Dilemma (PD), the Hawk-Dove (HD), and the
Stag-Hunt (SH) which is also called the Snowdrift Game or Chicken. For the sake of completeness, we briefly summarize
the main features of these games here; more detailed accounts
can be found in many places, for instance~\cite{axe84,poundstone92,skyrms04}. These games are all two-person, two-strategy, symmetric games with the payoff bi-matrix of Table~\ref{pbm}.
\begin{table}[hbt]
\begin{center}
{\normalsize
$
\begin{array}{c|cc}
 & C & D\\
\hline
C & (R,R) & (S,T)\\
D & (T,S) & (P,P)
\end{array}
$}
\end{center}
\caption{Generic payoff bi-matrix for the two-person, symmetric games
discussed in the text.\label{pbm}}
\end{table}
\noindent In this matrix, $R$ stands for the \textit{reward}
the two players receive if they
both cooperate ($C$), $P$ is the \textit{punishment} for bilateral defection ($D$), and $T$  is the
\textit{temptation}, i.e.~the payoff that a player receives if it defects, while the
other cooperates. In this case, the cooperator gets the \textit{sucker's payoff} $S$.
In the three games, the condition $2R > T + S$ is imposed so that
mutual cooperation is preferred over an equal probability of unilateral cooperation and defection.
For the PD, the payoff values are ordered numerically in the following way: $T > R > P > S$. 
Defection is always the best rational individual choice; 
$(D,D)$ is the unique NE and also an ESS~\cite{weibull95}.
Mutual cooperation  would be preferable but it is a strongly dominated strategy. 

In the Hawk-Dove game, the order of $P$ and $S$ is reversed yielding $T > R > S > P$. Thus, in the HD when both players defect they each get
the lowest payoff.
$(C,D)$ and $(D,C)$ are NE of the game in pure strategies, and there is
a third equilibrium in mixed strategies where strategy $D$ is played
with probability $p$, and strategy $C$ with probability $1-p$, where $p$ depends on the actual
payoff values.
The only ESS of the game is the mixed strategy,
while the two pure NE are not ESSs~\cite{weibull95}.
The dilemma in this game is caused by ``greed'', i.e.~players have a strong incentive
to ``bully'' their opponent by playing $D$, which is harmful for both parties if the outcome produced happens to be $(D,D)$.

In the Stag-Hunt, the ordering is $R > T > P > S$, which means that mutual cooperation $(C,C)$ is the best outcome,
Pareto-superior, and a NE.  However, there is a second NE equilibrium where both players defect
$(D,D$) which is inferior from the Pareto domination point of view, but it is less risky since it is safer
to play $D$ when there is doubt about which equilibrium should be selected. From a NE standpoint, however,
they are equivalent. Here the dilemma is represented by the fact that the
socially preferable coordinated equilibrium $(C,C)$ might be missed for ``fear'' that the other player
will play $D$ instead. 
There is a third
mixed-strategy NE in the game, but it is commonly dismissed because of its
inefficiency and also because it is not an ESS~\cite{weibull95}.

\section{Network Types}
\label{nets} 
 
 For our purposes here, a network will be represented as an undirected graph $G(V,E)$, where the
 set of vertices $V$ represents the agents, while the set of edges $E$ represents their symmetric interactions. The
 population size $N$ is the cardinality of $V$. A neighbor of an agent $i$ is any other agent $j$ such that there is an edge $\{ij\}  \in E$.
The cardinality of the set  of neighbors $V_i$  of player $i$ is the degree $k_i$ of vertex $i \in V$. The average
degree of the network will be called $\bar k$. An important quantity that will be used in the following
is the \textit{degree distribution function} (DDF) of a graph $P(k)$ which gives the probability that a given node has exactly
$k$ neighbors.

\begin{figure} [!ht]
\begin{center}
\begin{tabular}{cc}
	\mbox{\includegraphics[width=4cm]{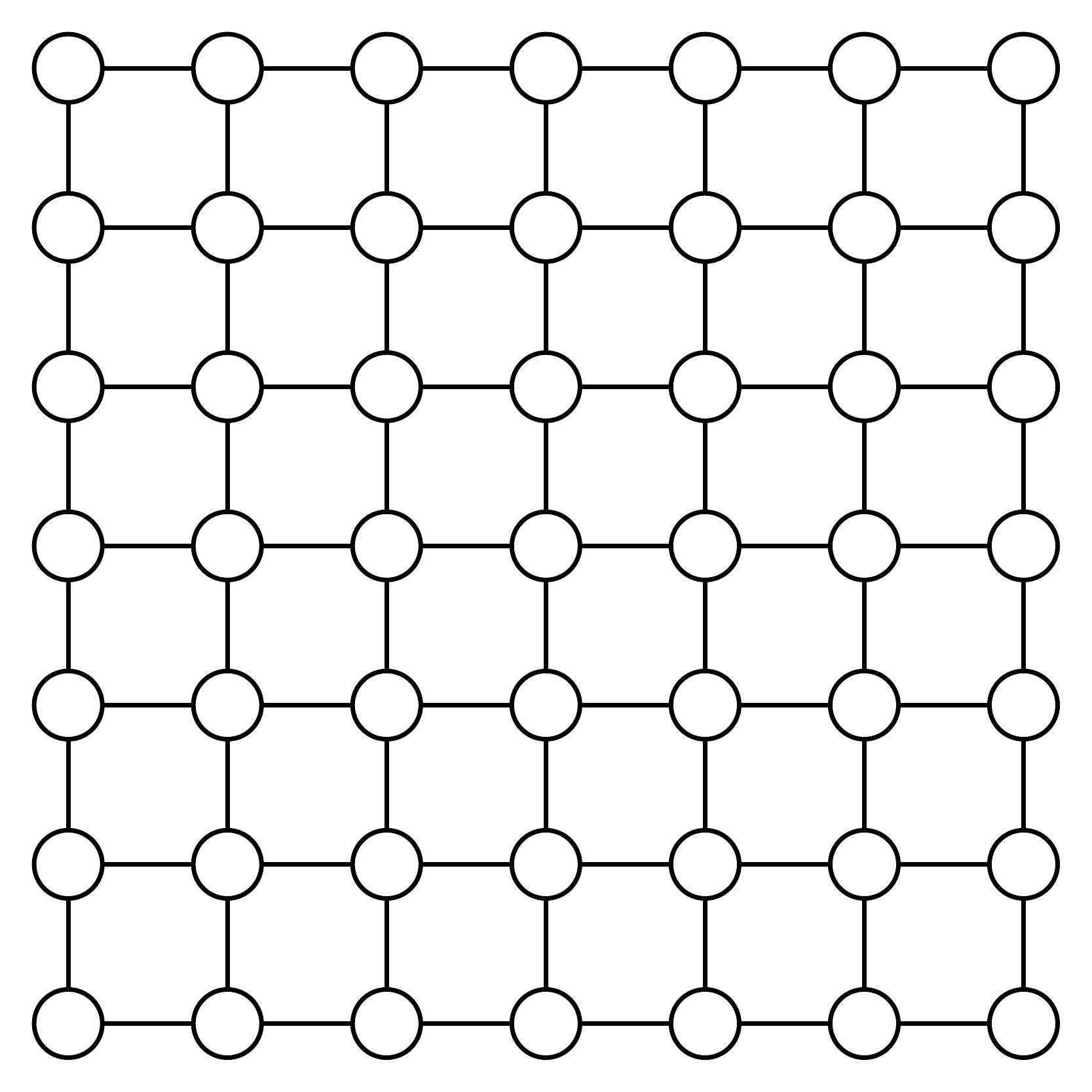}} \protect &
	\mbox{\includegraphics[width=6cm]{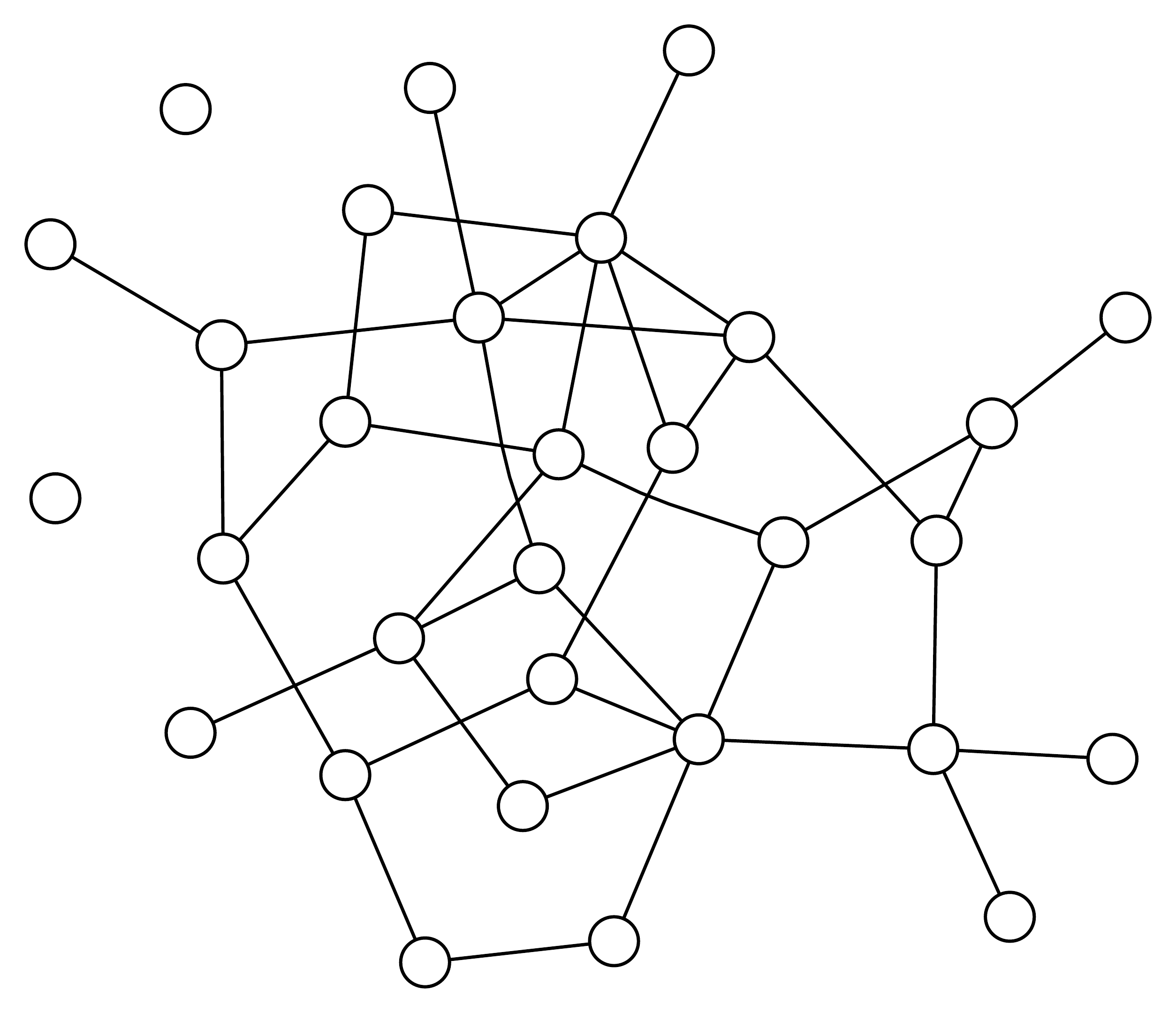}}\protect\\	
	\vspace*{0.5cm}(a)   &  (b) \\
	\multicolumn{2}{c}{\mbox{\includegraphics[width=6cm]{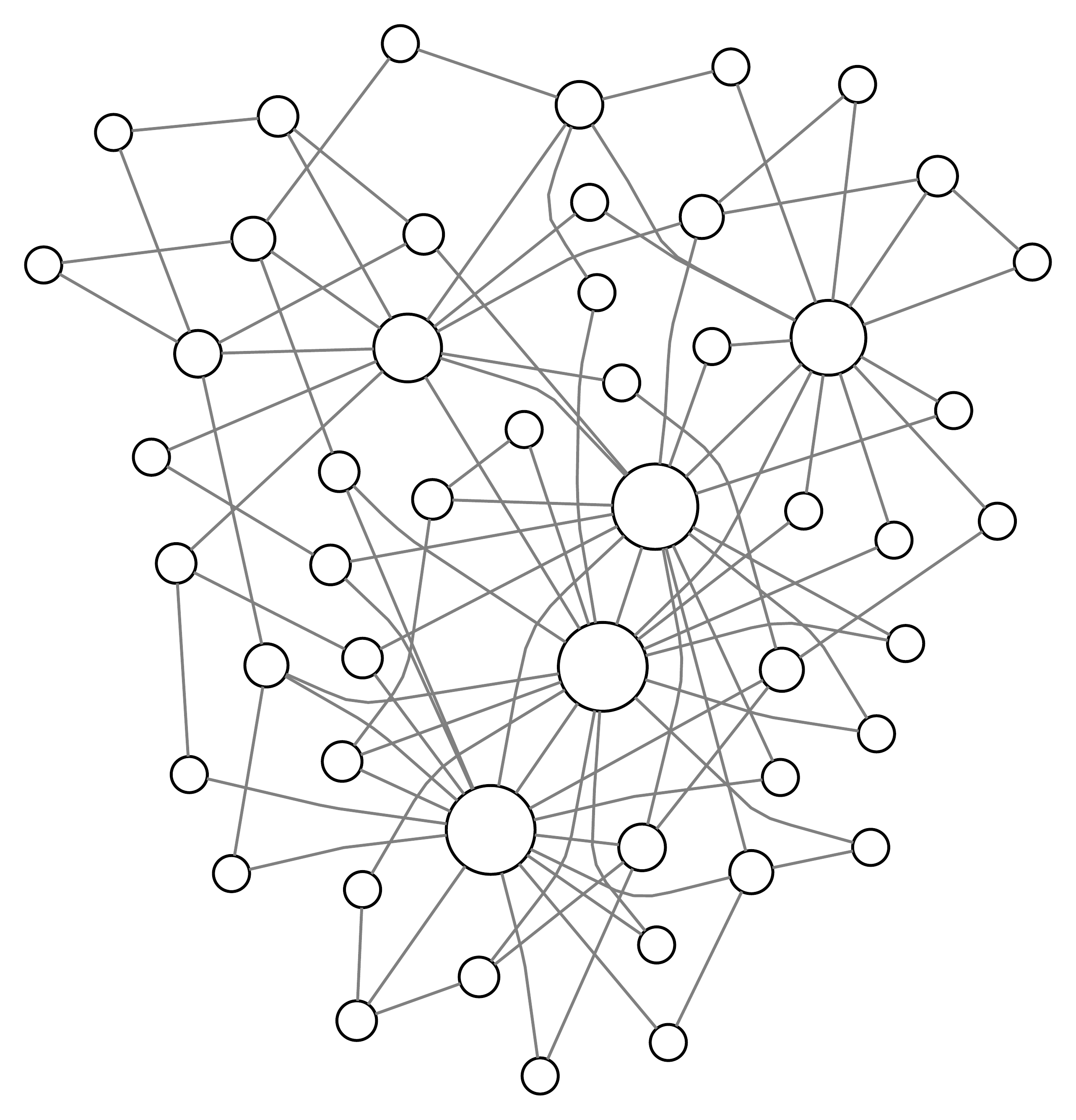}} \protect}\\
	\multicolumn{2}{c}{(c)} \\
\end{tabular}
\caption{A regular lattice (a), a random graph (b), and a scale-free graph (c). In (c) the nodes are shown with
 a size proportional to their number of neighbors.
\label{net_types}}
\end{center}
\end{figure}

To expose the technical problems and their solution,
we shall investigate three main graph population structures: regular lattices, random graphs, and scale-free graphs.
These graph types represent the typical extreme situations studied in the literature.
Regular lattices are examples of degree-homogeneous networks, i.e.~all the nodes have the same number of
neighbors; they have been studied from the EGT point of view in~\cite{nowakmay92,nowaketal94,nowak-sig-00,hauer-doeb-2004}, among others. In random graphs the degree fluctuates around the mean $\bar k$ but the fluctuations are small, of the order
of the standard deviation of the associated Poisson distribution. The situation can thus be described in mean-field
terms and is similar to the standard setting of EGT, where the large mixing population can be seen as a completely
connected graph. On the other hand, scale-free graphs
are typical examples of degree-heterogeneous graphs as the degree distribution is broad (see below). 
For the sake of illustration, examples of these three population network types are shown in Fig.~\ref{net_types}. For random and
scale-free graphs only one among the many possible realizations is shown, of course.

Recent work~\cite{newman-03} has shown that scale-free and other small-world graphs
are structurally and statistically much closer to actual social and biological networks and are thus an
interesting case to study. Evolutionary games on scale-free and other small-world networks have been investigated, among others, in
~\cite{social-pd-kup-01,santos-pach-05,tom-luth-giac-06,santos-pach-06}. 
Another interesting result for evolutionary games on networks has been recently obtained by 
Ohtsuki et al.~\cite{ohtsuki-et-al}.
In this study the authors present a simple rule for the evolution of cooperation on graphs based
on cost/benefit ratios and the number of neighbors of a given individual. 
This result is closely related to the subject matter of the present work but its application in the present
context will be the subject of further study. Our main goal is to consider
the global influence of network structure on the dynamics using a particular strategy update rule.
A further step toward
real social structures has been taken in~\cite{luthi-pest-tom-physa08}, where some evolutionary games are studied using
model social networks and an actual coauthorship network.

The DDF of  a regular graph is
 a normalized delta function centered at the constant degree $k$ of the graph. 
 Random graphs, which behave similar to panmictic populations, are constructed according to the standard
 Erd\"os--R\'enyi~\cite{bollobas-random} model: every possible edge among the $N$ vertices is present with probability $p$ or is absent with
 probabililty $1-p$. The DDF of such a random graph is Poissonian for $N \rightarrow  \infty$. Thus most
 vertices have degrees close to the mean value $\bar k$. In contrast, DDFs for complex networks
 in general have a longer tail to the right, which means that nodes with many neighbors may appear with
 non-negligible probability. An extreme example are scale-free networks in which the DDF is
 a power-law
  $P(k) \propto k^{- \gamma} $. Scale-free networks have been empirically found in many fields of technology,
  society, and science~\cite{newman-03}.
 To build scale-free networks, we use the model proposed by Barab\'asi and Albert ~\cite{alb-baraba-02}. In this model, networks are grown incrementally
starting with a small clique 
of $m_0$ nodes.
At each successive time step a new node is added such that its $m \le m_0$ edges link it to
$m$ nodes already present in the graph. It is
assumed that the probability $p$ that a new node will be connected to node $i$ depends on
the current degree $k_i$ of the latter. This is called the \textit{preferential attachment} rule. 
The probability $p(k_i)$ of node $i$ to be chosen is given by $p(k_i) = {k_i}/ \sum_{j} k_j,$ 
where the sum is over all nodes already in the graph.
The model evolves into a stationary network
with power-law probability distribution for the vertex degree $P(k) \sim k^{-\gamma}$, with
$\gamma\sim 3$.

\section{Replicator Dynamics in Networks}
\label{rd}

The local dynamics of a player $i$ only depends on its own strategy and on the strategies of the $k_i$ players in its neighborhood $V_i$.
Let us call $\pi_{ij}$ the payoff player $i$ receives when interacting with neighbor $j$. Let $M$ be the payoff
matrix corresponding to the row player. Since the games used here are symmetric the corresponding payoff
matrix of the column player is simply $M^T$, the transpose of $M$. For example, from table~\ref{pbm} of section
~\ref{games} one has:

\begin{center}
\begin{tabular}{cc}
\mbox{$ M =
 \begin{pmatrix}
R & S   \\
T &  P 
\end{pmatrix}$}, &~
\mbox{$ M^T =
 \begin{pmatrix}
R & T  \\
S &  P 
\end{pmatrix}$},
\end{tabular}
\end{center}

where suitable numerical values must be replaced for $R,S,T,P$.

This payoff $\pi_{ij}$ of the row player is now defined as
$$
\pi_{ij}(t) =  s_i(t)\; M\; s_{j}^T(t),
$$
\noindent where $s_i(t)$ and $s_j^T(t)$ are, respectively, row and column  vectors representing the 
players' mixed strategies i.e., the
probability distributions over the rows or columns played by $i$ and $j$ at time $t$. A pure strategy is the particular case
in which only one row or column is chosen.
The quantity
$$
 \widehat{\Pi}_i(t) = \sum _{j \in V_i}\pi_{ij}(t)
$$
\noindent is the \textit{accumulated payoff} collected by player $i$ at time step $t$, whereas the quantity
$\overline{\Pi}_i(t) = \frac{1}{k_i} \widehat{\Pi}_i(t)$ is his \textit{average payoff}. 

Accumulated payoff seems more logical in degree-heterogeneous networks such as scale-free
graphs since it
reflects the very fact that players may have different numbers of neighbors in the network.
Average payoff, on the other hand, smooths out the possible differences although it might
be justified in terms of number of interactions that a player may sustain in a given time.
For instance, an individual with many connections is likely to interact less often with each of its neighbors than another that has a lower number of connections. Also, if there is a cost to maintain a relationship,
average payoff will roughly capture this fact, while it will be hidden if one uses
accumulated payoff. On the other hand, if in a network some individuals happen to have many
more connections than the majority, this also means that they have somehow been able to
establish and maintain them; maybe this is a result of better social skills, more opportunities or for
other reasons but it is something that is commonly observed on actual social networks.
Because of this, most recent papers dealing with evolutionary games on networks have used accumulated 
payoff~\cite{social-pd-kup-01,santos-pach-05,santos-pach-06,santos-biol-06,luthi-pest-tom-physa08}, 
and this is the  main reason why we have focused on the technical problems that this
may cause in degree-heterogeneous networks.

The rule according to which agents update their strategies is the conventional RD.
The RD rule in networks aims at maximal consistency with the original
evolutionary game theory equations and is the same as proposed by~\cite{hauer-doeb-2004}.  It is assumed that the probability of switching strategy is
a monotonic increasing function $\phi$ of the payoff difference~\cite{weibull95,hofb-sigm-book-98}. To update the strategy
of player $i$, another player $j$ is first drawn uniformly at random from $i$'s neighborhood $V_i$. 
Then, strategy $s_i$ is replaced by $s_j$ with probability
\begin{equation}
 p_i = \phi(\Pi_j - \Pi_i),
 \label{repl_dyn_eq0}
\end{equation}
Where $\Pi$ may stand either for the above defined accumulated $\widehat{\Pi}$ or average $\overline{\Pi}$ payoffs,
or for the modified accumulated payoff $\widetilde{\Pi}$ to be defined below.
The major difference with standard replicator dynamics is that two-person encounters between players are only possible among neighbors, instead of being drawn from the whole population.
Other commonly used strategy update rules include imitating the best in the neighborhood, or replicating in proportion to the payoff, meaning that each individual $i$ reproduces with probability $p_i = \pi_i / \sum_j \pi_j$, where $pi_i$
is $i$'s payoff and the sum is over all $i's$ neighbors~\cite{hauer-doeb-2004}. However, in the present work we do not
examine these alternative rules.
Finally, contrary to~\cite{santos-pach-05}, we use asynchronous dynamics in the simulations presented here.
More precisely, we use the discrete update dynamics that makes the least assumption about the update sequence: the next node to be updated is chosen at random with uniform probability and with replacement.
This asynchronous update is analogous to the one used by Hauert et al.~\cite{hauer-doeb-2004}.
It corresponds to a binomial distribution of the updating probability and is a good approximation of a continuous-time Poisson process. We believe that asynchronous update dynamics are more likely in a system of independently interacting agents that may act at different and possibly uncorrelated times.
Furthermore, it has been shown that asynchronous updating may give rise to steadier quasi-equilibrium states by eliminating artificial effects caused by the nature of perfect synchronicity~\cite{tom-luth-pest-07}.
Nevertheless, in this work, we have checked that synchronous update of the agents' strategies does not qualitatively change the conclusions.

\subsection{Payoff Invariance}
\label{payoff-inv}

In standard evolutionary game theory one finds that
replicator dynamics is invariant
under positive affine transformations of payoffs with merely
a possible change of time scale~\cite{weibull95}.
Unfortunately, on degree-heterogenous networks, this
assumption is not satisfied when combining replicator
dynamics together with accumulated payoff.
This can be seen as follows.
Let $p_i$ in Eq. \ref{repl_dyn_eq0} be given by the following expression, as defined by Santos and Pacheco~\cite{santos-pach-05},
\begin{eqnarray}
p_i = \phi(\Pi_j -\Pi_i)  =
\begin{cases} \dfrac{\Pi_j - \Pi_i}{d_{M}k_>} & \textrm{if $\Pi_j
- \Pi_i > 0$}\\\\
0 & \textrm{otherwise,}
\end{cases}
\label{repl_dyn_eq1}
\end{eqnarray}
with $d_{M} = max\{T, R, P, S\} - min\{T, R, P, S\}$,
$k_> = max\{k_i, k_j\}$, and $\Pi_i$ (respectively $\Pi_j$) the aggregated payoff of a player $i$ (respectively $j$).
If we set $\Pi_x = \widehat{\Pi}_x$ for all $x \in V$ and now apply a positive affine transformation of the
payoff matrix, this leads to the new aggregated payoff
$$
\widehat{\Pi}^{'}_i = \sum_{j \in V_i}{\pi_{ij}^{'}} = \sum_{j \in V_i}{(\alpha\pi_{ij} + \beta)} = \alpha\sum_{j \in V_i}{\pi_{ij} + \sum_{j \in V_i}\beta} = \alpha \widehat{\Pi}_i + \beta k_i
$$
with $\alpha > 0, \beta \in \mathbb{R}$
and hence
\begin{eqnarray*}
\phi(\widehat{\Pi}_{j}' - \widehat{\Pi}_{i}') & = & (\alpha\widehat{\Pi}_j + \beta k_j -
\alpha\widehat{\Pi}_i - \beta k_i)/(\alpha d_{M}k_>)\\
& = &  \phi(\widehat{\Pi}_{j} - \widehat{\Pi}_{i}) + \beta (k_j - k_i)/(\alpha
d_{M}k_>).
\end{eqnarray*}
One can clearly see that using accumulated payoff does not
lead to an invariance of the replicator dynamics under
shifts of the payoff matrix. \\
As for the average payoff, although it respects the
replicator dynamics invariance under positive affine
transformation,
it prevents nodes with many edges to have potentially a
higher payoff than those with only a few links.
Furthermore, nodes are extremely vulnerable to defecting
neighbors with just one link.\\
Thus, we propose here a third definition for a player's payoff that retains the advantages of the accumulated and average payoff definitions
without their drawbacks.
Let $\pi_\gamma$ denote the guaranteed minimum payoff a player can obtain in a one-shot two-person game.
This is what a player would at least receive were he to attempt to maximize his minimum payoff.
For example in the PD, a player could choose to play $C$ with the risk of obtaining
the lowest payoff $S$ were its opponent to play $D$. However, by opting for strategy $D$ a player
would maximize its minimum payoff thus guaranteeing itself at least $\pi_\gamma = P > S$
no matter what its opponent's strategy might be.
In the HD game we have $\pi_\gamma = S$, for this time the payoff ordering is $T > R > S > P$ and a player
needs only to play $C$ to receive at least payoff $S$. Finally, in the SH game, $\pi_\gamma = P$.
We can now define a player $i$'s aggregated payoff as being
$\widetilde{\Pi}_i = \sum_{j \in V_i}{(\pi_{ij} - \pi_\gamma)}.$
Intuitively, it can be viewed as the difference between the payoff an individual collects and the minimum payoff it would get by ``playing it safe''. 
Our modified payoff $\widetilde{\Pi}$ has the advantage of leaving the RD invariant with respect to a positive affine transformation of the payoff matrix both on degree-homogeneous and heterogeneous graphs
while still allowing the degree distribution of the network to have a strong impact on the dynamics of the game.
Indeed, a player placed on a highly connected node of a graph can benefit from its numerous interactions
which enables it to potentially collect a high payoff.
However, these same players run the risk of totaling a much lower score than a player with only a few links.
One can notice that on degree-homogeneous graphs such as lattices or complete graphs, using accumulated,
average, or the new aggregated payoff definition yields the same results.
The proof of the RD invariance under positive affine transformation of the payoff matrix when using this new payoff definition is straightforward:
\begin{eqnarray*}
\phi(\widetilde{\Pi}_{j}' - \widetilde{\Pi}_{i}') & = & \frac{1}{\alpha
d_{M}k_>}\left(\sum_{k \in V_j}{\bigl((\alpha\pi_{jk} +
\beta) - (\alpha\pi_\gamma + \beta)}\bigr)\right.\\
& & \left.- \sum_{k \in V_i}{\bigl((\alpha\pi_{ik} + \beta)
- (\alpha\pi_\gamma + \beta)}\bigr)\right)\\
& = & \frac{1}{\alpha d_{M}k_>}\left(\alpha\sum_{k \in
V_j}{(\pi_{jk} - \pi_\gamma)}\right.\\
& & \left.- \alpha\sum_{k \in
V_i}{(\pi_{ik} - \pi_\gamma)}\right)\\
& = & (\widetilde{\Pi}_j - \widetilde{\Pi}_i)/(d_{M}k_>)\\
 & = & \phi(\widetilde{\Pi}_{j} - \widetilde{\Pi}_{i}).
\end{eqnarray*}

\subsection{Modified Replicator Dynamics}
\label{mrd}

Let us turn our attention once again to the replicator
dynamics rule (Eq.\ref{repl_dyn_eq1}). Dividing the payoff difference between players $j$ and $i$
by $d_{M}k_>$ might seem reasonable
at first since it does ensure that $\phi$ is a probability, i.e.\ has a value between 0 and 1.
Nevertheless, we don't find it to be the adequate division
to do for subtle reasons.
To illustrate our point, let us focus on the following
particular case and use the accumulated payoff to
simplify the explanation.
\begin{figure*} [!ht]
\begin{center}
\begin{tabular}{ccc}
	\vspace*{0.2cm}\mbox{\includegraphics[width=3.5cm,
height=3.5cm]{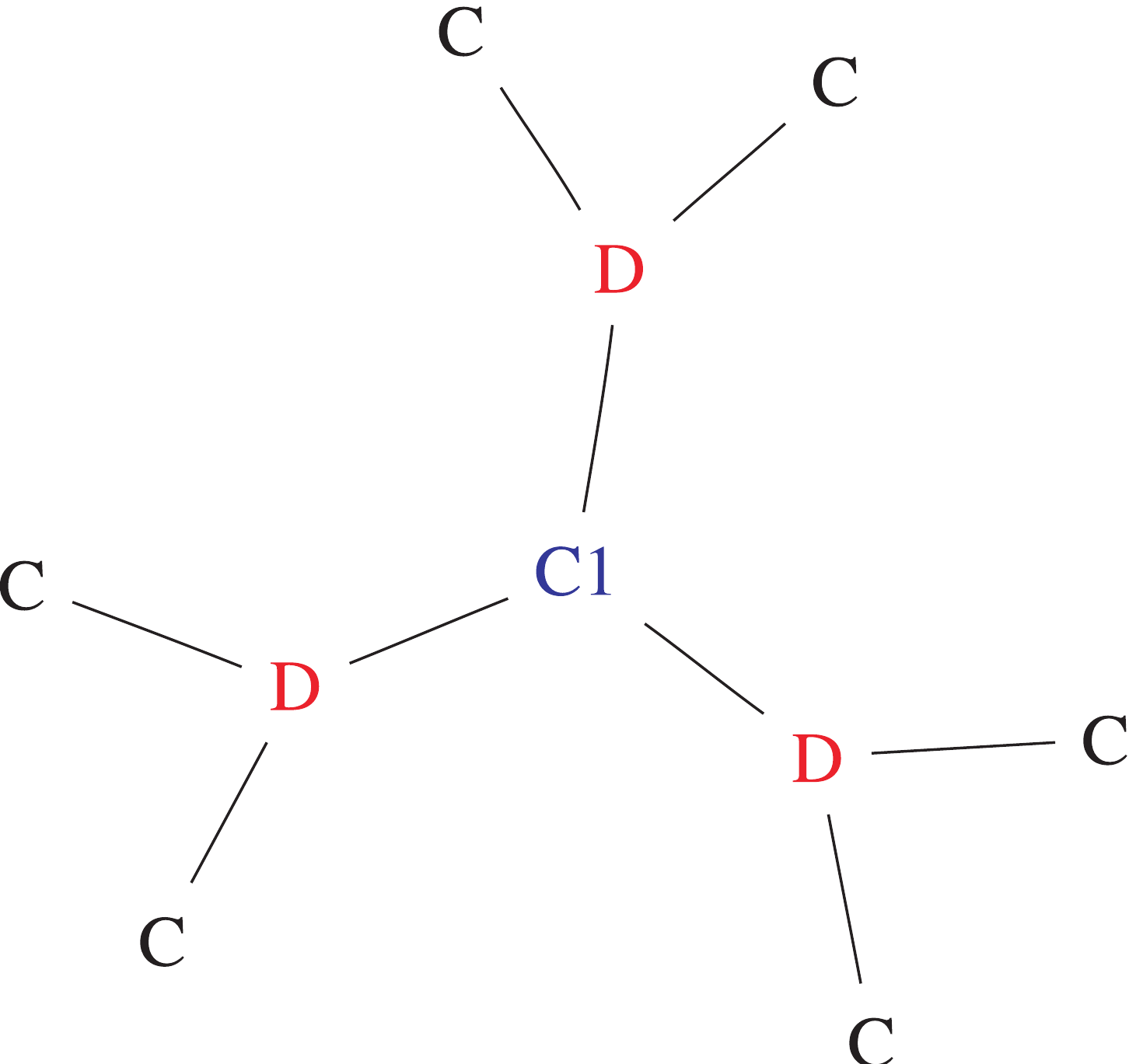}} \protect &
	\hspace*{1cm} &
	\mbox{\includegraphics[width=3.5cm,
height=3.5cm]{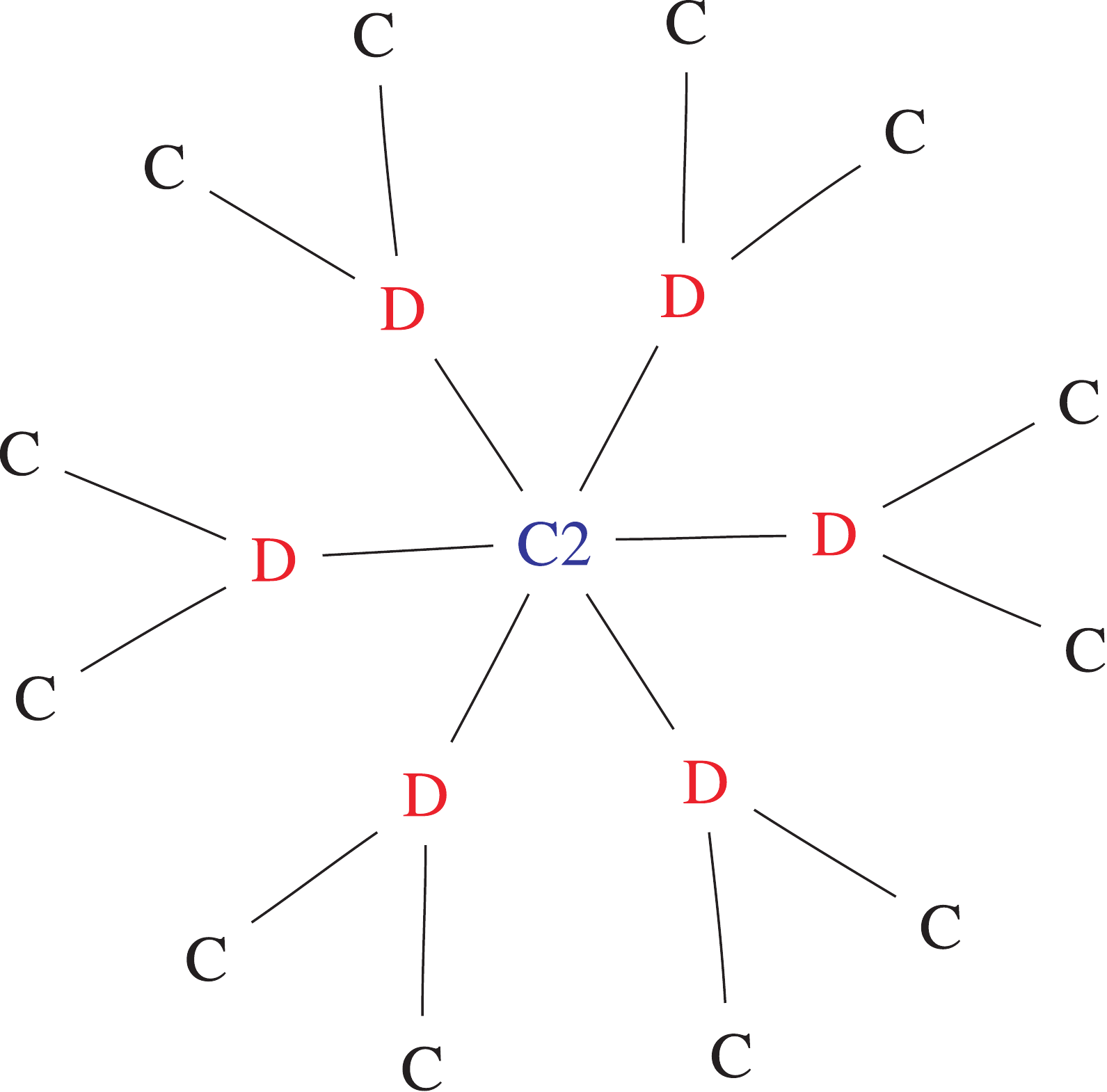}} \protect\\
	(a) & \hspace*{1cm} &  (b)
\end{tabular}
\caption{Example\label{example}}
\end{center}
\end{figure*}

On the one side, Fig.~\ref{example}~(a) shows a cooperator
$C1$ surrounded by
three defectors each having three cooperating neighbors.
Using the replicator dynamics as defined in
Eq.~\ref{repl_dyn_eq1},
the probability cooperator $C1$ would turn into a defector,
given that it is selected to be updated,
is equal to
\begin{eqnarray*}
\phi(\widehat{\Pi}_j - \widehat{\Pi}_{C1}) & = & (\widehat{\Pi}_j  - \widehat{\Pi}_{C1})/(d_{M}k_>)\\
& = & (3T - 3S)/(3d_{M})\\
& = & (T - S)/d_{M},
\end{eqnarray*}
and this no matter which defecting neighbor $j$ is chosen
since they all have the same payoff.
On the other side, the central cooperator $C2$ in
Fig.~\ref{example}~(b) would adopt strategy $D$ with probability
\begin{eqnarray*}
\phi(\widehat{\Pi}_j - \widehat{\Pi}_{C2}) & = & (\widehat{\Pi}_j  - \widehat{\Pi}_{C2})/(d_{M}k_>)\\
& = & (3T  - 6S)/6d_{M}\\
& = & (T - 2S)/2d_{M},
\end{eqnarray*}
a value that is once again independent of the selected
neighbor $j$.
Now, if $T > 0$ and $\phi(\widehat{\Pi}_j - \widehat{\Pi}_{C1}),\phi(\widehat{\Pi}_j - \widehat{\Pi}_{C2}) > 0$, then $C2$ has a bigger chance of having its strategy unaltered than $C1$ does.
This last statement seems awkward since in our opinion, the
fact of being surrounded by twice as many defectors
as $C1$ (with all the $D$-neighbors being equally strong),
should have a negative impact on cooperator $C2$,
making it difficult for it to maintain its strategy.
To make the situation even more evident, let us also suppose
$S = 0$. In this case, a cooperator surrounded by
an infinite number of $D$-neighbors, who in turn all have a
finite number of neighbors,
would have a zero probability of changing strategy, which is
counter-intuitive.
Therefore, and with all the previous arguments in mind, we
adjust Eq.~\ref{repl_dyn_eq1} to define another replicator dynamics function namely
\begin{eqnarray}
\phi(\Pi_j - \Pi_i)  =
\begin{cases} \dfrac{\Pi_j - \Pi_i}{\Pi_{j,\textrm{max}} - \Pi_{i,\textrm{min}}} & \textrm{if $\Pi_j - \Pi_i > 0$}\\\\
0 & \textrm{otherwise,}
\end{cases}
\label{repl_dyn_eq2}
\end{eqnarray}
where $\Pi_{x,\textrm{max}}$ (resp.\ $\Pi_{x,\textrm{min}}$) is the maximum (resp.\ minimum) payoff a player $x$ can get.
If $\pi_{x,\textrm{max}}$ and $\pi_{x,\textrm{min}}$ denote player $x$'s maximum and minimum payoffs in a two-player one-shot game ($\pi_{x,\textrm{max}} = max\{T,R,P,S\}$ and $\pi_{x,\textrm{min}} = min\{T,R,P,S\}$ for the dilemmas studied here), we have:
\begin{itemize}
\item $\Pi_{x, \textrm{max}} = \pi_{x,\textrm{max}}$ and $\Pi_{x, \textrm{min}} = \pi_{x,{\textrm{min}}}$ for average payoff;

\item $\Pi_{x, \textrm{max}} = k_x\pi_{x,\textrm{max}}$ and $\Pi_{x, \textrm{min}} = k_x\pi_{x,\textrm{min}}$ for accumulated payoff;

\item $\Pi_{x, \textrm{max}} = k_x(\pi_{x,\textrm{max}} - \pi_{x,\gamma})$ and $\Pi_{x, \textrm{min}} = k_x(\pi_{x,\textrm{min}} - \pi_{x,\gamma})$ for the new payoff scheme.
\end{itemize}

Finally, one can easily verify that using $\Pi_i = \widetilde{\Pi}_i$ as the aggregated payoff of a player $i$ leaves equation Eq.~\ref{repl_dyn_eq2} invariant with respect to a positive affine transformation of the payoff matrix.

\begin{figure} [!ht]
\begin{center}
\includegraphics[width=14.5cm,bb=0 0 1404.365 1299.3849]{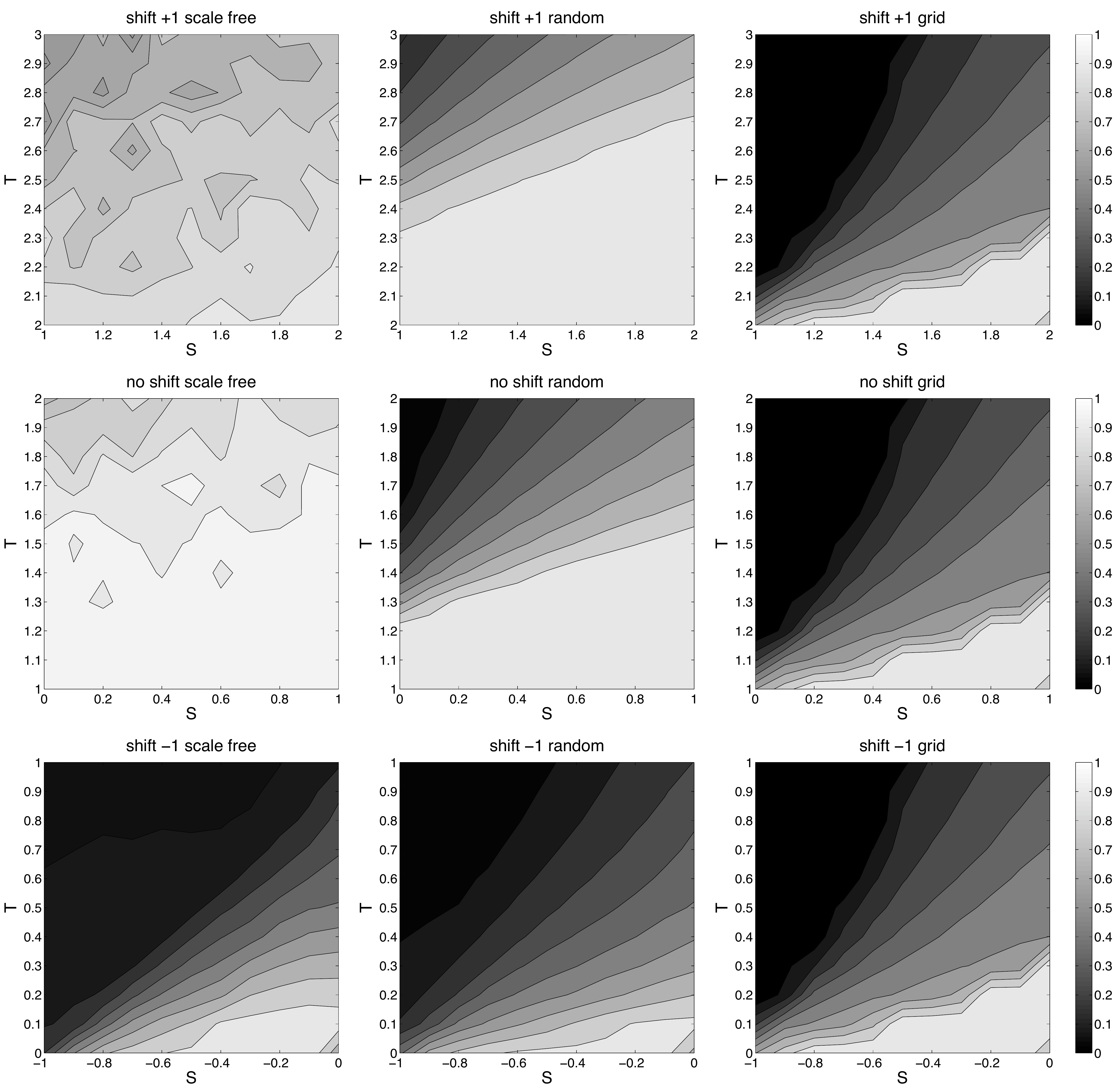}
\caption{Amount of cooperation in the HD game using accumulated payoff on three different network types in three different game spaces (see text). Lighter areas mean more cooperation than darker ones (see scale on the right side). Left column: scale free; Middle column: random graph; Right column: grid. Upper row: $2 \le T \le 3$, $R=2$, $1 \le S \le 2$, $P = 1$; Middle row: $1 \le T \le 2$, $R=1$, $0 \le S \le 1$, $P = 0$; Bottom row: $0 \le T \le 1$, $R=0$, $-1 \le S \le 0$, $P = -1$\label{shifts-acc}}
\end{center}
\end{figure}

\section{Numerical Simulations}
\label{ns}

\begin{figure} [!ht]
\begin{center}
\begin{tabular}{ccc}
	\vspace*{-0.2cm}
	\mbox{\includegraphics[width=5.5cm] {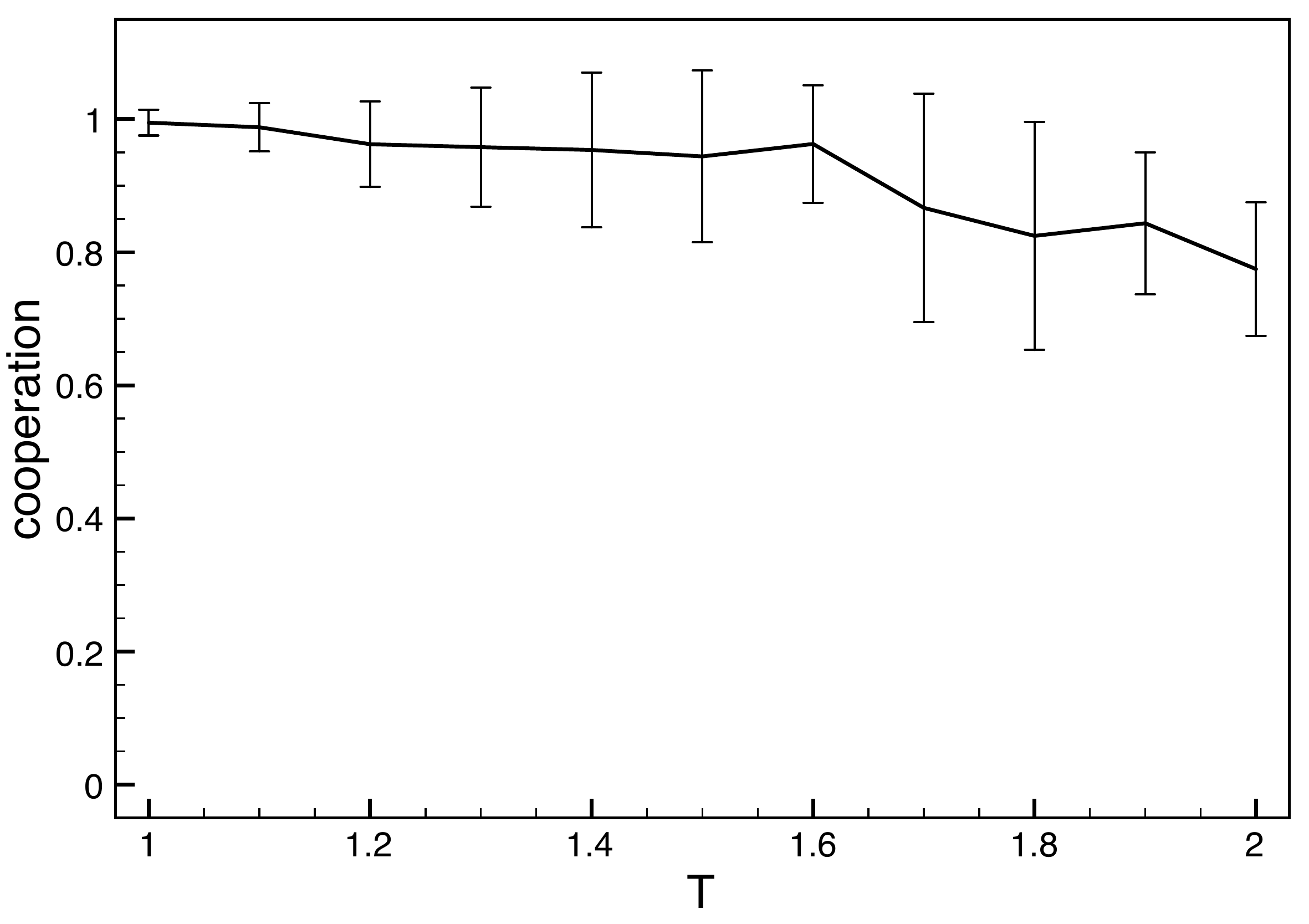}} \protect &&
	\mbox{\includegraphics[width=5.5cm] {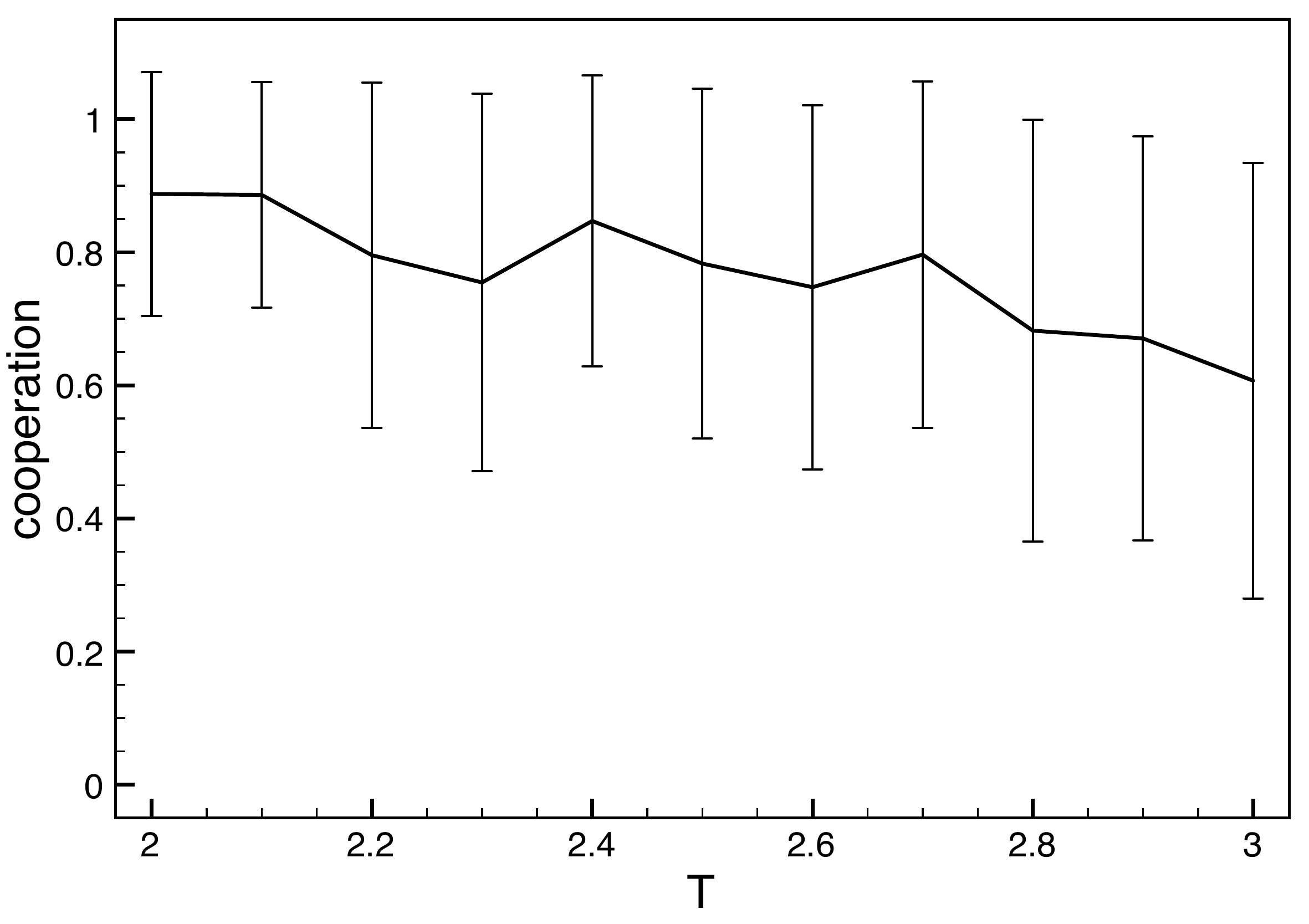}} \protect\vspace*{0.2cm}\\
	(a) & \hspace*{.1cm} &  (b)
\end{tabular}
\caption{Standard deviation for the HD using accumulated payoff on scale-free networks for two different game spaces. (a)  $1 \le T \le 2$, $R=1$, $S=0.1$, $P = 0$, (b)  $2 \le T \le 3$, $R=2$, $S=1.1$, $P = 1$.
Note that (a) is a cut at $S=0.1$ of the middle image in the leftmost column of Fig.~\ref{shifts-acc}, while (b)
represents a cut of the topmost image in the leftmost column of Fig.~\ref{shifts-acc} at $S=1.1$.\label{accumulated_deviation}}
\end{center}
\end{figure}
We have simulated the PD, HD and SH described in Sect.~\ref{games} on regular lattices,  Erd\"os--R\'enyi random graphs and Barab\'asi--Albert scale-free graphs, all three of which were presented in Sect.~\ref{nets}. Furthermore, in each case, we test the three payoff schemes discussed in Sect.~\ref{rd}.
\begin{figure} [!ht]
\begin{center}
\includegraphics[width=14.5cm,bb=0 0 1404.365 449.3849]{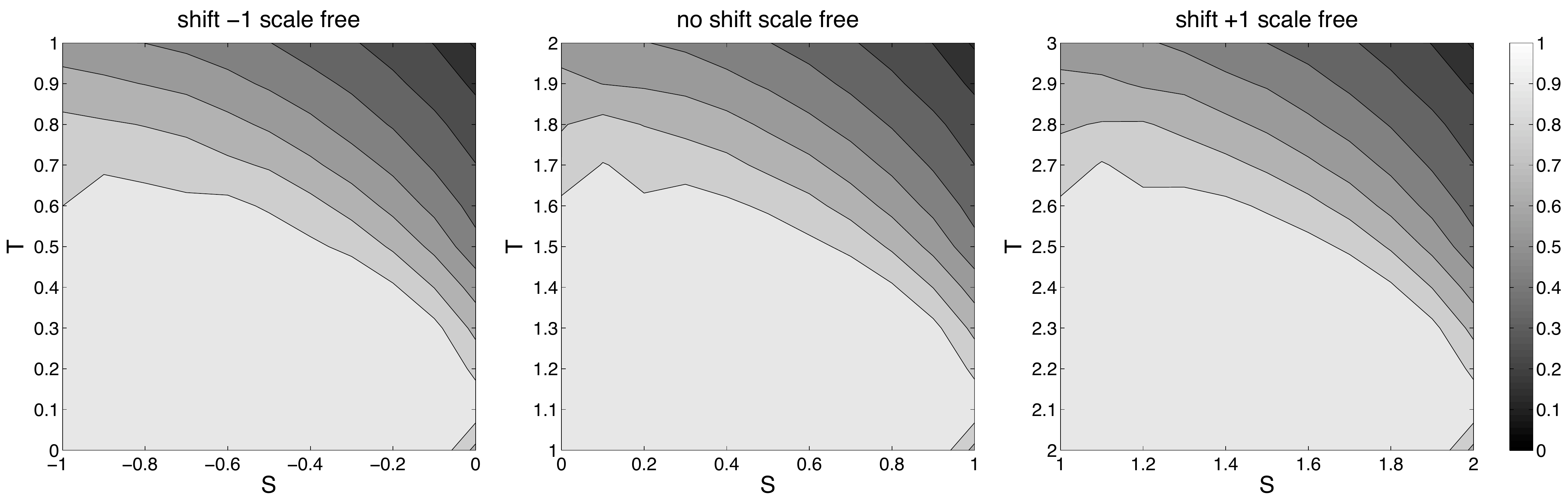}
\caption{Levels of cooperation in the HD game using the new aggregated payoff $\widetilde{\Pi}$ on scale-free graphs in three different game spaces (see text). Left: $2 \le T \le 3$, $R=2$, $1 \le S \le 2$, $P = 1$; Middle: $1 \le T \le 2$, $R=1$, $0 \le S \le 1$, $P = 0$; Right: $0 \le T \le 1$, $R=0$, $-1 \le S \le 0$, $P = -1$.\label{shifts_inv}}
\end{center}
\end{figure}
The networks used are all of size $N=4900$ with an average degree $\overline{k} = 4$.
The regular lattices are two-dimensional with periodic boundary conditions, and
the neighborhood of an individual comprises the four closest individuals in the north, east, south, and west directions.
The Erd\"os--R\'enyi random graphs were generated using connection probability $p=8.16\times10^{-4}$.
Finally, the Barab\'asi--Albert were constructed starting with a clique of $m_0=2$ nodes and at each time step the new incoming node has $m=2$ links.\\
For each game, we limit our study to the variation of only two parameters per game.
In the case of the PD, we  set $R=1$ and $S=0$,
and vary $1 \leq T \leq 2$ and $0 \leq P \leq 1$.
For the HD, we set $R=1$ and $P=0$ and the two parameters are $1 \leq T \leq 2$ and $0 \leq S \leq 1$.
Finally, in the SH, we decide to fix $R = 1$ and $S = 0$ and vary $0 \leq T \leq 1$ and $0 \leq P \leq T$.\\
We deliberately choose not to vary the same two parameters in all three games.
The reason we choose to set $T$ and $S$ in both the PD and the SH is
to simply provide natural bounds on the values to explore of  the remaining two parameters.
In the PD case, $P$ is limited between $R=1$ and $S=0$
in order to respect the ordering of the payoffs ($T>R>P>S$) and $T$'s upper bound is equal to 2  due to the
$2R > T+S$ constraint.
In the HD, setting $R=1$ and $P=0$ determines the range of $S$ (since this time $T>R>S>P$)
and gives an upper bound of 2 for $T$, again due to the $2R > T+S$ constraint.
Note however, that the only valid value pairs of $(T,S)$ are those that satisfy the latter constraint.
Finally, in the SH, both $T$ and $P$ range from $S$ to $R$.
Note that in this case, the only valid value pairs of $(T,P)$ are those that satisfy $T>P$.\\
It is important to realize that, when using our new aggregated payoff or the average payoff, even though we reduce our study to the variation of only two parameters per game, we are actually exploring the entire game space.
This is true owing to the invariance of Nash equilibria and replicator dynamics under positive affine transformations of the payoff matrix~\cite{weibull95}.
As we have shown earlier and as we will confirm numerically in the next section, this does not hold for the accumulated payoff.\\
Each network is randomly initialized with exactly 50\% cooperators and 50\% defectors.
In all cases, the parameters are varied between their two bounds by steps of $0.1$.
For each set of values, we carry out 50 runs of 15000 time steps each, using
a fresh graph realization in each run.
Cooperation level is averaged over the last 1000 time steps, well after the transient equilibration period.
In the figures that follow, each point is the result of averaging over 50 runs.
In the next two sections, in order to avoid overloading this document with figures, we shall focus each time on one of the three games, commenting on the other two along the way.

\subsection{Payoff Shift}
\label{pay-shift}
We have demonstrated that in theory, the use of accumulated payoff does not leave the RD invariant under positive affine transformations of the payoff matrix. However, one can wonder whether in practice, such shifts of the payoff matrix translate into significant differences in cooperation levels or are the changes just minor.

\begin{figure} [!ht]
\begin{center}
\includegraphics[width=14.5cm,bb=0 0 1404.365 900]{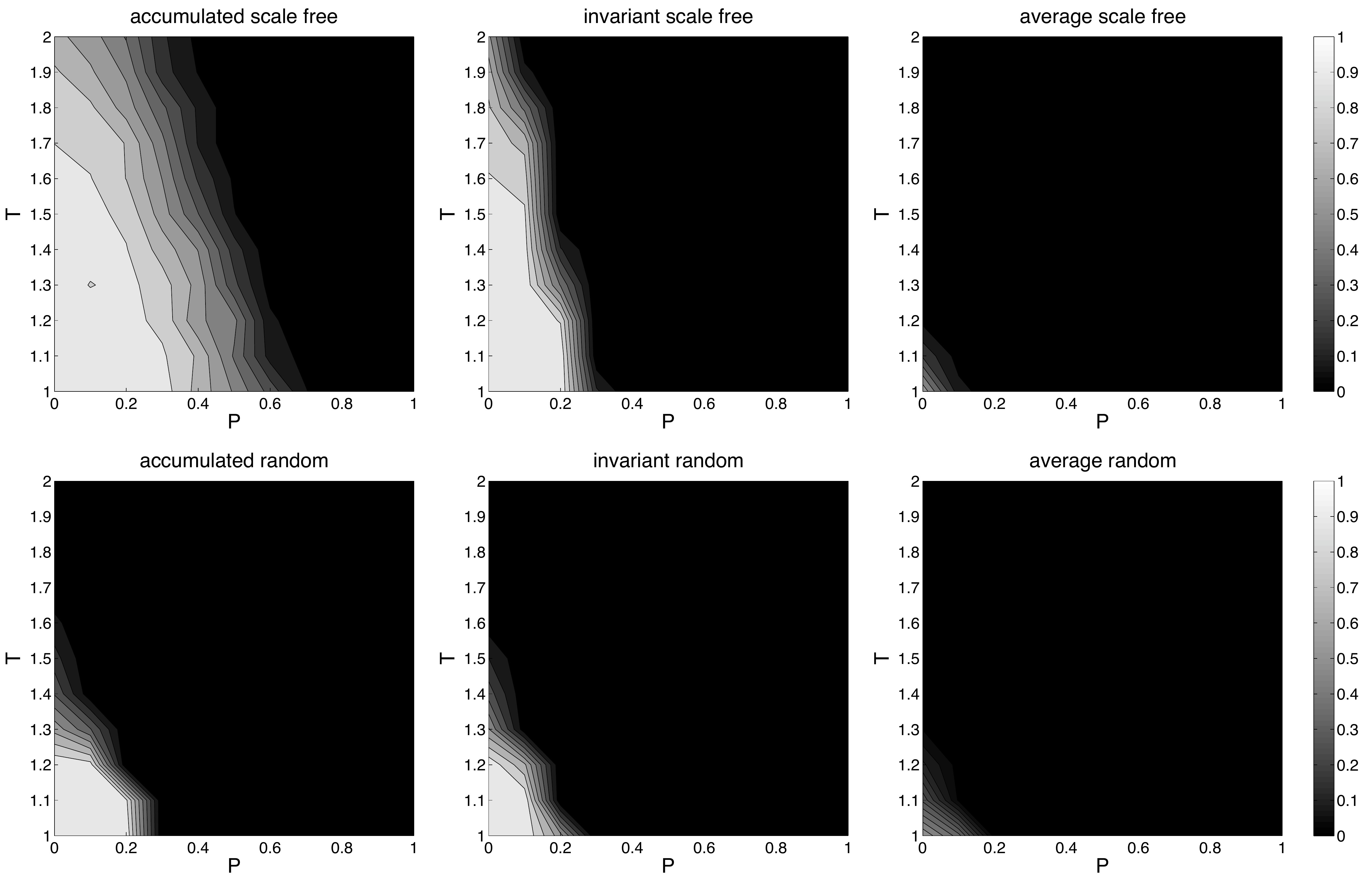}
\caption{Levels of cooperation in the PD game space using three different payoff schemes and two different network types. Left column: Accumulated Payoff; Middle column: New Aggregated Payoff; Right column: Average Payoff. Upper row: Scale free graph; Bottom row: Random graph.
Game space: $1\le T \le 2$, $R=1$, $0 \le P \le 1$, $S = 0$.\label{PD}}
\end{center}
\end{figure}

Figure~\ref{shifts-acc} depicts the implications of a slight positive and negative shift of the HD payoff matrix. As one can clearly see, the cooperation levels encountered are notably different before and after the shift. As a matter of fact, when comparing between network types, scale-free graphs seem to do less well in terms of cooperation than regular grids with a shift of $-1$, and not really better than random graphs with a
shift of $+1$.   
Thus, one must be extremely cautious when focusing on a rescaled form of the payoff matrix, affirming that such a re-scaling can be done without loss of generality, for this is far from true when dealing with accumulated payoff.\\
The noisy aspect of the top two figures of the leftmost column of Fig.~\ref{shifts-acc} has caught our attention. It is essentially due to the very high standard deviation values we find in the given settings (see Fig.~\ref{accumulated_deviation}).
 This observation is even more pronounced with a shift of $+1$. This shows that replicator dynamics becomes relatively unstable when using straight accumulated payoff.\\

We have run simulations using our payoff $\widetilde{\Pi}$, on all three network types in order to numerically validate the invariance of the RD with this payoff scheme. However, to save space, we only show here the results obtained on scale-free graphs which are the networks that generated the biggest differences in the accumulated payoff case (see Fig.~\ref{shifts-acc}, leftmost colummn). As one can see in Fig.~\ref{shifts_inv}, using $\widetilde{\Pi}$ does indeed leave the RD invariant with respect to a shift of the payoff matrix. There are minor differences between the figures, but these are simply due to statistical sampling and roundoff errors.
Finally, a shift of the payoff matrix has, as expected, no influence at all on the general outcome when using the average payoff. We point out that the same observations can also be made for the PD and SH cases (not shown here).

\subsection{Payoff and Network Influence on Cooperation}
\label{res}

\begin{figure} [!ht]
\begin{center}
\includegraphics[width=14.5cm,bb=0 0 1404.365 900]{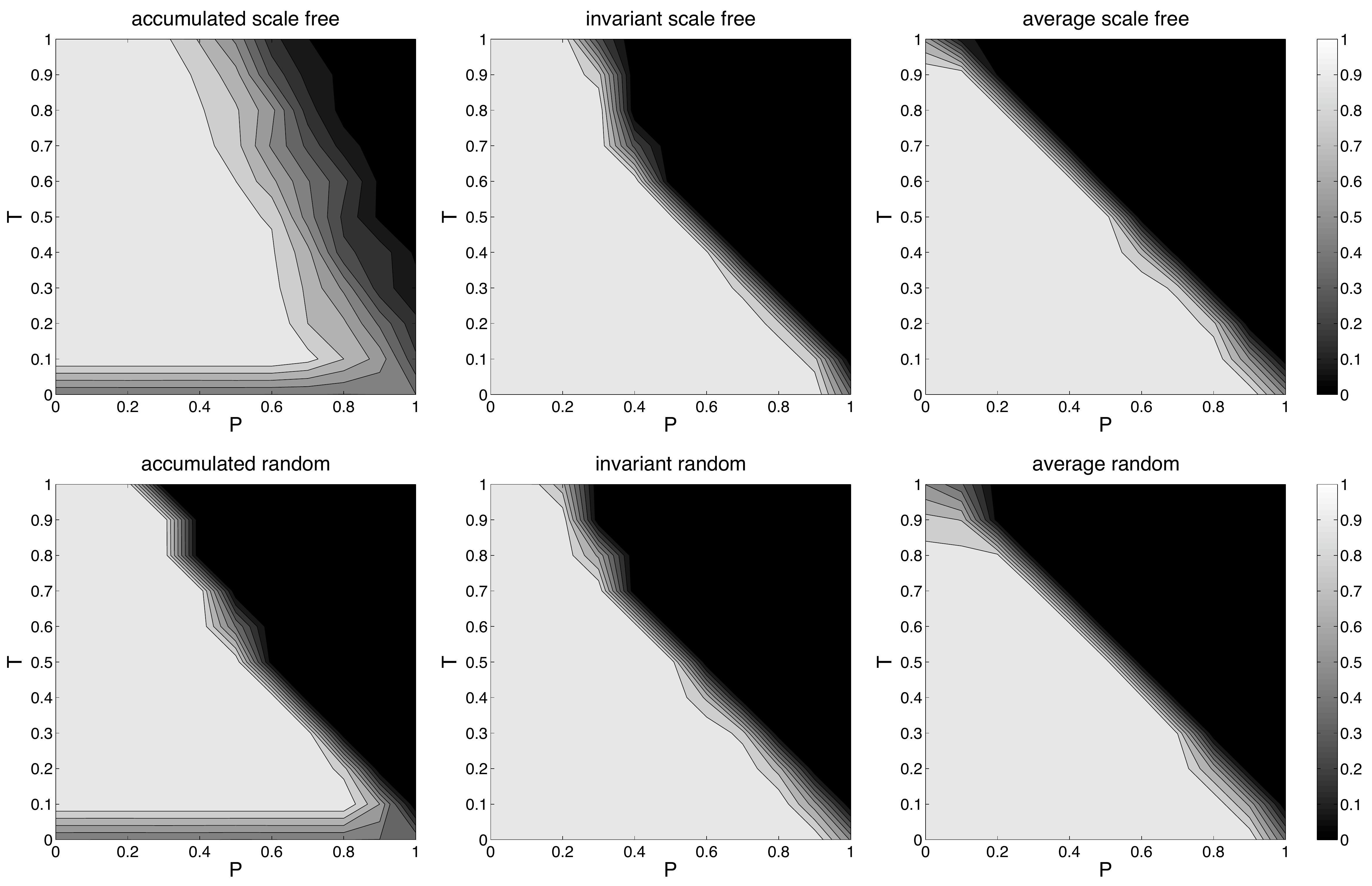}
\caption{Cooperation levels for the SH game space using three different payoff schemes and two different network types. Left column: Accumulated Payoff; Middle column: New Aggregated Payoff; Right column: Average Payoff. Upper row: Scale free graph; Bottom row: Random graph.
Game space: $R=1$, $0\le T \le 1$, $0 \le P \le 1$, $S = 0$.
Note that the meaningful game space is the upper left triangle, i.e.~when $T \geq P$.
\label{SH}}
\end{center}
\end{figure}

In this section we report results on global average cooperation levels using the three payoff schemes
for two games on scale-free and random graphs.\\
Figure~\ref{PD} illustrates the cooperation levels reached for the PD game, in the $1\le T \le 2$, $R=1$, $0 \le P \le 1$, $S = 0$ game space, on a Barab\'asi--Albert scale-free and random graphs, and when using
each of the three different payoff schemes mentioned earlier, namely $\overline{\Pi}$, $\widetilde{\Pi}$ and $\widehat{\Pi}$.\\
We immediately notice that there is a significant parameter zone for which accumulated payoff (leftmost column) seems to drastically promote cooperation compared to average payoff (rightmost column).
This observation has already been highlighted in some previous work~\cite{tom-luth-pest-07,santos-biol-06}, although it was done for a reduced game space.
We nevertheless include it here to situate the results obtained using our adjusted payoff in this particular game space in comparison to those obtained using the two other extreme payoff schemes.
On both network types, $\widetilde{\Pi}$ (central column of Fig.~\ref{PD}) yields cooperation levels somewhat like those obtained with accumulated payoff but to a lesser degree.
This is especially striking on scale-free graphs (upper row of Fig.~\ref{PD}). However, we again point out
that the situation shown in the image of the upper left corner of Fig.~\ref{PD} would change dramatically
under a payoff shift, as discussed in Sect.~\ref{pay-shift} for the HD game.
The same can be observed for the HD and SH games (see Fig.~\ref{SH} for the SH case).
On regular lattices, there are as expected no differences whatsoever between the use of $\widetilde{\Pi}$ over $\widehat{\Pi}$ or $\overline{\Pi}$ due to the degree homogeneity of this type of network (not shown).

\begin{figure} [!ht]
\begin{center}
\includegraphics[width=8cm]{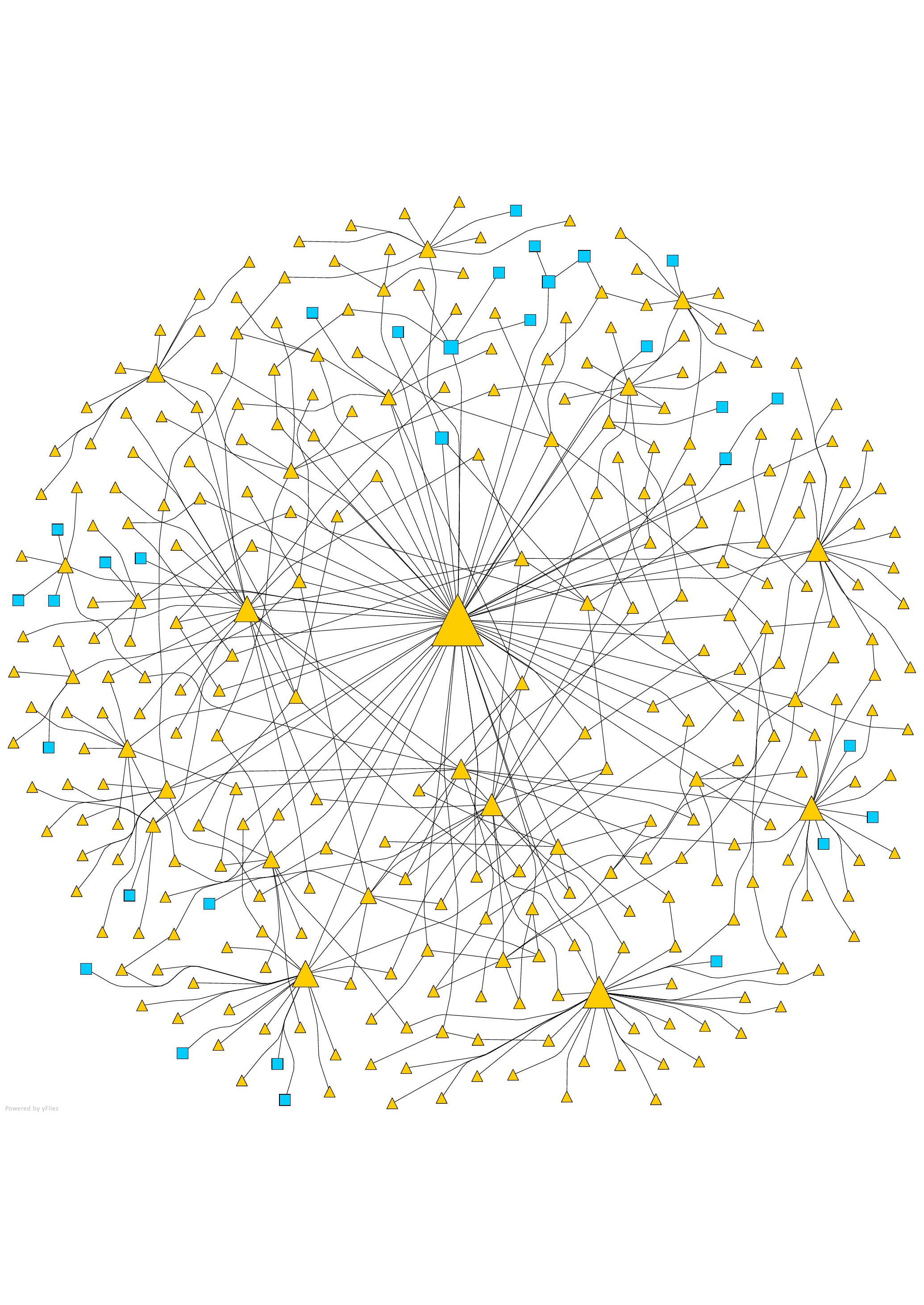}
\caption{A cluster with a majority of cooperators (triangles) with many links to a central cooperator. Symbol size is proportional to degree. Links to other nodes of the network have been
suppressed for clarity.\label{cluster}}
\end{center}
\end{figure}

The primary goals of this work are to highlight the non-invariance of the RD under affine transformations of the payoff matrix when using accumulated payoff, and to propose an alternative payoff scheme without this drawback.
How does the network structure influence overall cooperation levels when this latter payoff is chosen? 
Looking at the middle column of figures~\ref{PD} and~\ref{SH}, we observe that degree non-homogeneity enhances cooperation.
The relatively clear separation in the game space between strongly cooperative regimes and entirely defective ones in the middle column of Fig.~\ref{SH}, which refers to the SH game, can be explained by the existence of the two ESSs in pure strategies in this case.
Similarly, the large transition phase from full cooperation to full defect states in the HD (middle image of
Fig.~\ref{shifts_inv}) is due to the fact that the only ESS for this game is a mixed strategy.\\
Cooperation may establish and remain stable in networks thanks to the formation of clusters of
cooperators, which are tightly bound groups of players. In the scale-free case this is easier for,
as soon as a highly connected node becomes a cooperator, if a certain number of its neighbors
are cooperators as well, chances are that all neighbors will imitate the central cooperator, which
is earning a high payoff thanks to the number of acquaintances it has. An example of such a
cluster is shown in Fig.~\ref{cluster} for the PD. A similar phenomenon has been found to underlie cooperation
in real social networks~\cite{luthi-pest-tom-physa08}.

In order to explore the dependence of the evolutionary processes on the network size, we have
performed simulations with two other graph sizes ($N=2450$ and $N=9800$) for the HD game.
To save space, we do not show the figures but cooperation results are qualitatively very similar to those 
shown here for $N=4900$. We have also simulated populations with two different initial percentages
of randomly distributed cooperators: $30\%$ and $70\%$; again, there are no qualitative differences
with the $50$-$50$ case shown here.

\section{Conclusions}
\label{concl}

Standard RD assumes infinite mixing populations of playing agents. Actual and simulated populations
are necessarily of finite size and show a network of ties among agents that is not random, as postulated
by the theory. In this work we have taken the population finiteness for granted and we have focused
on the graph inhomogeneity aspects of the problem. It is a well known fact that agent clustering may
provide the conditions for increased cooperation levels in games such as those studied here. However,
up to now, only regular structures such as grids had been studied in detail, with the exception of a few investigations that have dealt with small-world population structures of various kinds 
~\cite{social-pd-kup-01,santos-pach-05,tom-luth-giac-06,ohtsuki-et-al,luthi-pest-tom-physa08}. But most
have used an accumulated payoff scheme that makes no difference in regular graphs, but in the other cases, it does not leave the RD invariant with respect to affine transformations of the payoff matrix,
which is required by evolutionary game
theory. This gives rise to results that are not generalizable to the whole game space. The alternative of
using average payoff respects invariance but is much less realistic in degree-inhomogeneous networks
that are the rule in society. Here we have proposed
a new payoff scheme that correctly accounts for the degree inhomogeneity of the underlying population
graph and, at the same time, is invariant with respect to these linear transformations. Using this
scheme, we have shown that, on complex networks, cooperation may reach levels far above
what would be predicted by the standard theory for extended regions of the game's parameter
space. The emergence of cooperation is essentially due to the progressive colonization by cooperators
of highly connected clusters in which linked cooperators that earn a high payoff mutually
protect themselves from exploiting defectors. This phenomenon had already been observed
to a lesser extent in populations structured as regular grids but it is obviously stronger for scale-free graphs,
where there exist a sizable number of highly connected individuals and it is the same effect that
underlies cooperation in actual social networks.
This observation alone may account for observed
increased levels of cooperation in society without having to take into account other factors such as
reputation, belonging to a recognizable group, or repeated interactions giving rise to complex reciprocating strategies, although these factors also play a role in the emergence of cooperation.

\section*{Acknowledgments} E. Pestelacci and M. Tomassini gratefully acknowledge financial support by the Swiss National Science Foundation under contract 200021-111816/1.

\bibliographystyle{elsart-num}
\bibliography{games}

\begin{thebibliography}{10}
\expandafter\ifx\csname url\endcsname\relax
  \def\url#1{\texttt{#1}}\fi
\expandafter\ifx\csname urlprefix\endcsname\relax\def\urlprefix{URL }\fi
\expandafter\ifx\csname href\endcsname\relax
  \def\href#1#2{#2} \def\path#1{#1}\fi

\bibitem{maynard82}
J.~M. Smith, Evolution and the Theory of Games, Cambridge University Press,
  1982.

\bibitem{weibull95}
J.~W. Weibull, Evolutionary Game Theory, MIT Press, Boston, MA, 1995.

\bibitem{hofb-sigm-book-98}
J.~Hofbauer, K.~Sigmund, Evolutionary Games and Population Dynamics, Cambridge
  University Press, Cambridge, UK, 1998.

\bibitem{Myerson}
R.~B. Myerson, Game Theory: Analysis of Conflict, Harvard University Press,
  Cambridge, MA, 1991.

\bibitem{taylor-jonker}
P.~Taylor, L.~Jonker, Evolutionary stable strategies and game dynamics,
  Mathematical Biosciences 16 (1978) 76--83.

\bibitem{fogeletal97}
D.~B. Fogel, G.~B. Fogel, P.~C. Andrews, On the instability of evolutionary
  stable states, BioSystems 44 (1997) 135--152.

\bibitem{fogeletal98}
G.~B. Fogel, P.~C. Andrews, D.~B. Fogel, On the instability of evolutionary
  stable states in small populations, Ecological Modeling 109 (1998) 283--294.

\bibitem{ficici-pollack-07}
S.~G. Ficici, J.~B. Pollack, Evolutionary dynamics of finite populations in
  games with polymorphic fitness-equilibria, Journal of Theoretical Biology 247
  (2007) 426--441.

\bibitem{nowak-et-al-finite-04}
M.~A. Nowak, A.~Sasaki, C.~Taylor, D.~Fudenberg, Emergence of cooperation and
  evolutionary stability in finite populations, Nature 428 (2004) 646--650.

\bibitem{ficici-pollack-05}
S.~Ficici, O.~Melnik, J.~B. Pollack, A game-theoretic and dynamical systems
  analysis of selection methods in coevolution, IEEE Transactions on
  Evolutionary Computation 9~(6) (2005) 580--602.

\bibitem{axe84}
R.~Axelrod, The Evolution of Cooperation, Basic Books, Inc., New York, 1984.

\bibitem{nowakmay92}
M.~A. Nowak, R.~M. May, Evolutionary games and spatial chaos, Nature 359 (1992)
  826--829.

\bibitem{watts99}
D.~J. Watts, Small worlds: The Dynamics of Networks between Order and
  Randomness, Princeton University Press, Princeton NJ, 1999.

\bibitem{newman-03}
M.~E.~J. Newman, The structure and function of complex networks, {SIAM} Review
  45 (2003) 167--256.

\bibitem{social-pd-kup-01}
G.~Abramson, M.~Kuperman, Social games in a social network, Phys. Rev. E 63
  (2001) 030901.

\bibitem{santos-pach-05}
F.~C. Santos, J.~M. Pacheco, Scale-free networks provide a unifying framework
  for the emergence of cooperation, Phys. Rev. Lett. 95 (2005) 098104.

\bibitem{tom-luth-giac-06}
M.~Tomassini, L.~Luthi, M.~Giacobini, Hawks and doves on small-world networks,
  Phys. Rev. E 73 (2006) 016132.

\bibitem{luthi-pest-tom-physa08}
L.~Luthi, E.~Pestelacci, M.~Tomassini, Cooperation and community structure in
  social networks, Physica A 387 (2008) 955--966.

\bibitem{bollobas-random}
B.~Bollob\'as, Random Graphs, Academic Press, New York, 2001, 2nd ed.

\bibitem{samuel97}
L.~Samuelson, Evolutionary Games and Equilibrium Selection, MIT Press,
  Cambridge, MA, 1997.

\bibitem{poundstone92}
W.~Poundstone, The Prisoner's Dilemma, Doubleday, New York, 1992.

\bibitem{skyrms04}
B.~Skyrms, The Stag Hunt and the Evolution of Social Structure, {Cambridge
  University Press}, Cambridge, 2004.

\bibitem{nowaketal94}
M.~A. Nowak, S.~Bonhoeffer, R.~M. May, Spatial games and the maintenance of
  cooperation, Proc. Nat. Acad. Sci. USA 91 (1994) 4877--4881.

\bibitem{nowak-sig-00}
M.~A. Nowak, K.~Sigmund, Games on grids, in: U.~Dieckmann, R.~Law, J.~A.~J.
  Metz (Eds.), The Geometry of Ecological Interactions: Simplifying Spatial
  Complexity, Cambridge University Press, Cambridge, UK, 2000, pp. 135--150.

\bibitem{hauer-doeb-2004}
C.~Hauert, M.~Doebeli, Spatial structure often inhibits the evolution of
  cooperation in the snowdrift game, Nature 428 (2004) 643--646.

\bibitem{santos-pach-06}
F.~C. Santos, J.~M. Pacheco, T.~Lenaerts, Evolutionary dynamics of social
  dilemmas in structured heterogeneous populations, Proc. Natl. Acad. Sci. USA
  103 (2006) 3490--3494.

\bibitem{ohtsuki-et-al}
H.~Ohtsuki, C.~Hauert, E.~Lieberman, M.~A. Nowak, A simple rule for the
  evolution of cooperation on graphs and social networks, Nature 441~(7092)
  (2006) 502--505.
\newblock \href {http://dx.doi.org/http://dx.doi.org/10.1038/nature04605}
  {\path{doi:http://dx.doi.org/10.1038/nature04605}}.

\bibitem{alb-baraba-02}
R.~Albert, A.-L. Barabasi, Statistical mechanics of complex networks, Reviews
  of Modern Physics 74 (2002) 47--97.

\bibitem{santos-biol-06}
F.~C. Santos, J.~M. Pacheco, A new route to the evolution of cooperation,
  Journal of Theoretical Biology 19~(2) (2006) 726--733.

\bibitem{tom-luth-pest-07}
M.~Tomassini, E.~Pestelacci, L.~Luthi, Social dilemmas and cooperation in
  complex networks, Int: J. Mod. Phys. C 18~(7) (2007) 1173--1185.

\end{thebibliography}

\end{document}